\documentclass[
    pra,                
    aps,                 
    twocolumn,             
    groupedaddress,     
    longbibliography,       
    nofootinbib,           
    floatfix,
    10pt               
]{revtex4-2}

\usepackage[utf8]{inputenc}
\usepackage[T1]{fontenc}  
\usepackage{microtype}     
\microtypecontext{spacing=nonfrench}

\usepackage{amsmath}
\usepackage{amssymb} 
\usepackage{amsbsy}
\usepackage{amsfonts}
\usepackage{mathrsfs}
\usepackage{bm}

\usepackage{amsthm} 

\usepackage{graphicx} 
\usepackage{booktabs} 
\usepackage{multirow}
\usepackage{array}    
\usepackage[dvipsnames]{xcolor}
\usepackage{colortbl} 
\usepackage{makecell} 

\usepackage{textcomp}      

\usepackage{siunitx}
\usepackage[colorlinks=true,
            linkcolor=blue,       
            urlcolor=blue,        
            citecolor=blue,       
            anchorcolor=blue,
            breaklinks=true]{hyperref} 
\usepackage[ruled,vlined,linesnumbered]{algorithm2e} 

\usepackage[capitalize]{cleveref}

\DeclareMathOperator*{\argmin}{arg\,min}
\DeclareMathOperator*{\argmax}{arg\,max}

\begin{document}

\newcommand{\nudt}{{College of Computer Science and Technology, National University of Defense Technology, Changsha 410073, China}}
\author{Yuanqi Liu} \thanks{These authors contributed equally.}
\author{Weilei Zeng} \thanks{These authors contributed equally.}
\author{Junjie Wu}
\author{Lingling Lao} \email{laolinglingrolls@gmail.com}
\affiliation{\nudt}
\date{\today}

\title{Approximate maximum-likelihood decoding via truncated free energies}
  
\begin{abstract}

Maximum-likelihood decoding (MLD) achieves the minimum logical error rate of stabilizer codes under known i.i.d.\ Pauli noise, but its exact evaluation is \#P-hard.
Practical pipelines therefore approximate MLD by minimum-weight decoding (MWD), retaining only the lowest-weight recovery per syndrome and discarding the coset degeneracy. The minimum-weight search is in turn implemented by stochastic solvers.
We introduce approximate maximum-likelihood decoding (AMLD), a black-box framework that recycles the candidate samples discarded by stochastic inner decoders into a per-class truncated free-energy estimator. For every logical class represented in the candidate pool, the estimator is provably bounded below by the exact free energy and above by the empirical minimum weight. AMLD returns the logical class minimizing the estimated free energy with linear classical overhead.
In SA-based Ising-decoder benchmarks, AMLD closes up to $83\%$ of the MWD--MLD threshold gap across the toric and color codes under bit-flip and depolarizing noise.
The largest threshold improvement, from $17.28\%$ to $18.62\%$, occurs on the $6.6.6$ color code under depolarizing noise. We further demonstrate AMLD on the $[[144,12,12]]$ bivariate-bicycle code, whose bit-flip decoding problem has a hypergraph structure. This application requires neither matching-based enumeration nor code-specific tensor-network contraction. At $p=0.05$, AMLD reduces the logical error rate by $13\%$ relative to MWD evaluated on the same BP-OSD candidate pool. 
\end{abstract}

\maketitle
\section{Introduction}

Fault-tolerant quantum computation~\cite{shor1996fault,knill1998resilient,Preskill1998} based on quantum error correction~\cite{calderbankshor1996,Shor1995,Knill1997} relies on a classical decoder that infers recovery operations from the syndrome data~\cite{Kitaev2003,Fowler2012,Terhal2015}.
Under independent and identically distributed (i.i.d.) Pauli noise, maximum-likelihood decoding (MLD) selects the logical equivalence class minimizing the free energy derived from the partition function, a Boltzmann sum over all stabilizer-equivalent error operators in that class~\cite{Dennis2002,DuclosCianciPoulin2010}. MLD is the information-theoretically optimal decoder for stabilizer codes, but its exact evaluation is \#P-hard~\cite{IyerPoulin2015}. Practical pipelines therefore approximate MLD by minimum-weight decoding (MWD), which represents each class by its minimum-weight error configuration and thereby discards the coset degeneracy. MWD is itself NP-hard even for specific families of topological codes~\cite{mwd_np_hard2026}, and is implemented by stochastic heuristic solvers~\cite{Kirkpatrick1983,Hutter2014,takeuchi2023depolarizingising,takada2024colorising}. MWD and MLD can therefore yield different decoding thresholds for two-dimensional topological codes~\cite{Dennis2002,Bombin2012,BravyiSucharaVargo2014}, motivating systematic approximations to MLD.

Two lines of work on approximating MLD have emerged. Tensor-network contraction of the exact partition function approximates MLD but relies on code-specific geometry~\cite{BravyiSucharaVargo2014,Chubb2021,PiveteauChubbRenes2024}. Another strategy approximates the partition function by truncating it to a finite set of candidate errors. Matching-based examples include ensemble decoders such as Libra~\cite{Jones2024} and Harmony~\cite{ShuttyNewmanVillalonga2026}, which induce candidate diversity through per-shot perturbations of the decoder's assumed prior probabilities and then pool the class-resolved likelihoods. $K$-MWM decoding~\cite{lin_approximate_2025} instead deterministically enumerates the top-$K$ minimum-weight matchings to construct a Boltzmann-reweighted class probability mathematically equivalent to a truncated free-energy estimate. Its efficient exact top-$K$ enumeration applies to matching graphs; for correlated $X$--$Z$ errors, where the decoding problem is hypergraph-structured, it instead uses a heuristic based on separate matching searches in the $X$- and $Z$-subgraphs.

In this work, we show that candidate generation need not rely on matching-based enumeration: the configurations otherwise discarded by stochastic MWD solvers already encode enough free-energy information to partially close the MWD--MLD gap. Although a stochastic MWD solver does not generally sample from the Boltzmann distribution, its output concentrates on low-weight configurations, which dominate the partition function at the Nishimori temperature. 
We introduce approximate maximum-likelihood decoding (AMLD), a black-box framework that treats the per-syndrome candidate pool from a stochastic inner decoder as an empirical truncation of the per-class partition function and returns the class of lowest estimated free energy. Using only candidate configurations and their Pauli weights, AMLD decouples free-energy estimation from candidate generation. For any nonempty per-class candidate pool, we prove that the estimated free energy lies between the exact free energy and the empirical minimum weight. With one retained candidate per class, AMLD reduces to a minimum-weight decision over the sampled candidates; when every pool covers the full coset, it recovers exact MLD. Its advantage over MWD is most direct in the ground-state-dominated regime, where sampled ground-state multiplicities resolve cross-class minimum-weight ties that MWD would otherwise break uniformly.

We benchmark AMLD across three inner-decoder families spanning Ising Hamiltonian, linear-system, and message-passing formulations: the Ising decoder implemented by simulated annealing (SA)~\cite{Kirkpatrick1983}, the Random Windows (RW) decoder~\cite{dumer2017distance,pryadko2022qdistrnd}, and belief propagation with ordered-statistics decoding (BP-OSD)~\cite{Roffe2020}, whose candidate diversity is induced by per-qubit prior variation across repeated trials. With SA-based Ising decoding on the toric and color codes under bit-flip and depolarizing noise, AMLD closes up to $83\%$ of the MWD--MLD threshold gap. We further examine empirical trends in AMLD's advantage with the many-body interaction order $R$ and the logical-class count $|\mathcal{L}|$; these quantities are not varied independently of the code family and noise model. With RW, a stochastic linear-system inner decoder, AMLD and MWD yield statistically indistinguishable threshold estimates, showing no statistically significant change relative to the shared-pool MWD baseline. Using BP-OSD-generated candidates for the $[[144,12,12]]$ bivariate-bicycle code~\cite{Bravyi2024} under bit-flip noise at $p=0.05$, AMLD reduces the logical error rate by $13\%$ relative to MWD evaluated on the same candidate pool. The code has a hypergraph-structured decoding problem with $|\mathcal{L}|=4096$ logical classes. This benchmark demonstrates AMLD in the quantum LDPC regime without requiring matching-based enumeration or code-specific tensor-network contraction~\cite{Breuckmann2021,panteleev2020QuantumLDPCCodes,tillich2013QuantumLDPCCodes,zeng2019HigherDimensionalQuantumHypergraphProduct,zeng2020MinimalDistancesCertain}.

This paper is structured as follows. We first review the decoding framework and its statistical-mechanics mapping in Sec.~\ref{sec:preliminaries}. 
Then we formalize the AMLD framework and its degeneracy-resolution mechanism in Secs.~\ref{sec:algorithm} and \ref{sec:theory}.
The numerical results are presented in Sec.~\ref{sec:numerics}. We conclude the paper in Sec.~\ref{sec:discussion}.

\section{Decoding framework and statistical mechanics mapping}
\label{sec:preliminaries}

\subsection{Quantum error correction codes}
\label{sec:codes}

\textit{Toric code.}---The toric code~\cite{Kitaev2003} is defined on an $L \times L$ square lattice with periodic boundary conditions, placing $n = 2L^2$ data qubits on the lattice edges. Its stabilizer group is generated by vertex operators $A_v = \prod_{q \in \delta v} X_q$ and face operators $B_f = \prod_{q \in \partial f} Z_q$, each of weight four, yielding code parameters $[[2L^2, 2, L]]$. A single-qubit $X$ (or $Z$) error violates exactly two stabilizers of the opposite Pauli type, so the decoding problem reduces to matching pairs of violated stabilizers on a graph whose edges correspond to data qubits.

\textit{$6.6.6$ color code.}---The $6.6.6$ color code~\cite{Bombin2006} is defined on a hexagonal lattice with a triangular boundary and three-colorable faces, placing data qubits at the lattice vertices. Each face $f$ supports two stabilizer generators $S_f^X = \prod_{v \in \partial f} X_v$ and $S_f^Z = \prod_{v \in \partial f} Z_v$. The code encodes $k=1$ logical qubit with distance $d$ (the number of qubits along any boundary side), using $n = (3d^2+1)/4$ physical qubits for odd $d$. Each bulk qubit sits at a trivalent vertex incident to three faces, so a single-qubit $X$ (resp.\ $Z$) error violates the three $Z$-type (resp.\ $X$-type) stabilizers around that vertex, inducing a decoding hypergraph with 3-uniform hyperedges. Three hexagonal faces ($|\partial f|=6$) meet at each bulk vertex, hence the 6.6.6 nomenclature.

\textit{Bivariate bicycle codes.}---The bivariate bicycle (BB) codes~\cite{Bravyi2024} are quantum LDPC codes constructed from two polynomials $f(x,y), g(x,y) \in \mathbb{F}_2[x,y]/(x^{\ell_x}-1, y^{\ell_y}-1)$, with parity-check matrices $H^X = [f \mid g]$ and $H^Z = [g^T \mid f^T]$ in the cyclic-group representation. We use the $[[144, 12, 12]]$ code ($\ell_x=12, \ell_y=6$) for numerical evaluation. Under independent bit-flip noise, $X$ errors are labeled by their commutation with the $k$ $Z$-type logical operators alone, giving $|\mathcal{L}| = 2^k = 4096$ for the BP-OSD experiment in Sec.~\ref{sec:numerics}.

\subsection{Error decoding objectives}
\label{sec:stabilizer_codes}

We work within the stabilizer formalism over the $n$-qubit Pauli group~\cite{Gottesman1996,Calderbank1998}. For a stabilizer code encoding $k$ logical qubits with $r$ independent stabilizer generators, each syndrome measurement yields a binary vector $\boldsymbol{\gamma} \in \{0,1\}^r$. Because a single syndrome is consistent with exponentially many physical errors, the decoding objective is to identify the most likely logical equivalence class rather than any specific error representative~\cite{Poulin2006}.

Up to a global phase, each $n$-qubit Pauli error is represented by a binary vector $\boldsymbol{e}=(e^X\mid e^Z)\in\{0,1\}^{2n}$, referred to as its binary symplectic representation. The commutation relation of two Pauli operators $\boldsymbol{u}, \boldsymbol{v}$ is defined by their symplectic inner product, i.e., $\boldsymbol{u}\,\Omega\,\boldsymbol{v}^{T} = 0 (1) \pmod 2$ if they commute (anti-commute), with $\Omega = \bigl(\begin{smallmatrix}0&I_n\\I_n&0\end{smallmatrix}\bigr)$.
Let $S \in \{0,1\}^{r \times 2n}$ be the stabilizer (or parity-check) matrix, $L \in \{0,1\}^{2k \times 2n}$ be the logical generator matrix, and $D \in \{0,1\}^{r \times 2n}$ be the destabilizer matrix, with $D\,\Omega\,S^T = I_r \pmod 2$. These matrices form a symplectic basis, ensuring any error decomposes uniquely as
\begin{equation}
    \boldsymbol{e}(\boldsymbol{\gamma}, \boldsymbol{\alpha}, \ell) = \boldsymbol{\gamma} D + \boldsymbol{\alpha} S + \ell L \pmod 2,
    \label{eq:error-configuration}
\end{equation}
where $\boldsymbol{\gamma}D$ is a fixed pure error reproducing the syndrome, $(\boldsymbol{\gamma}D)\,\Omega\,S^T=\boldsymbol{\gamma}\pmod 2$, $\boldsymbol{\alpha}\in\{0,1\}^{r}$ parameterizes the stabilizer degrees of freedom within an equivalence class, and $\ell\in\mathcal{L}\subseteq\{0,1\}^{2k}$ is a binary logical-class label, while $\ell L\in\{0,1\}^{2n}$ represents the corresponding logical Pauli operator. For fixed $\ell$, the coset $C_\ell\equiv\{\boldsymbol{e}(\boldsymbol{\gamma},\boldsymbol{\alpha},\ell):\boldsymbol{\alpha}\in\{0,1\}^{r}\}$ collects all stabilizer-equivalent errors in class $\ell$. The label set $\mathcal{L}$ has cardinality $4^k$ for general Pauli noise and $2^k$ for purely bit-flip noise.

Under the i.i.d.\ Pauli channels considered here, $P(\boldsymbol{e})$ denotes the probability that the channel produces the error configuration $\boldsymbol{e}$. This probability depends on $\boldsymbol{e}$ only through its Pauli weight $w(\boldsymbol{e})$, the number of qubits acted on by a non-identity Pauli, via $-\ln P(\boldsymbol{e}) = \beta\, w(\boldsymbol{e}) + \mathrm{const}$, where $\beta$ is the channel-determined inverse temperature. MLD sums the probability mass over all stabilizer-equivalent errors within each class:
\begin{equation}
    \hat{\ell}_{\mathrm{MLD}} = \argmax_{\ell \in \mathcal{L}} \sum_{\boldsymbol{\alpha} \in \{0,1\}^{r}} P\bigl(\boldsymbol{e}(\boldsymbol{\gamma}, \boldsymbol{\alpha}, \ell)\bigr).
    \label{eq:MLD}
\end{equation}
MWD replaces the inner sum by its largest term:
\begin{equation}
    \hat{\ell}_{\mathrm{MWD}} = \argmax_{\ell \in \mathcal{L}}  \max_{\boldsymbol{\alpha} \in \{0,1\}^{r}} P\bigl(\boldsymbol{e}(\boldsymbol{\gamma}, \boldsymbol{\alpha}, \ell)\bigr),
    \label{eq:MWD}
\end{equation}
which is equivalent to selecting the class containing the lowest-weight representative. The discrepancy between Eq.~\eqref{eq:MLD} and Eq.~\eqref{eq:MWD} is purely entropic in origin. Its leading manifestation in the limit $p \to 0$ is the multiplicity of minimum-weight representatives: when several classes share the same minimum weight $w_{\min}$, MWD encounters ground-state degeneracy and breaks the tie with equal probability, whereas MLD resolves it through the full $\boldsymbol{\alpha}$-summation (see Sec.~\ref{sec:theory_degeneracy} for the analysis).

\subsection{Statistical mechanics mapping}
\label{sec:mapping}

As established in Sec.~\ref{sec:stabilizer_codes}, $-\ln P(\boldsymbol{e})$ is linear in $w(\boldsymbol{e})$. This maps error-weight minimization to a classical ground-state search and lets the class likelihoods in Eq.~\eqref{eq:MLD} be expressed as partition functions of disordered Ising models~\cite{sourlas_spin-glass_1989,Dennis2002,kovalev_spin_2015}. For independent bit-flip noise (pure $X$ errors), only the $Z$-type stabilizers detect errors. The relevant gauge freedom in Eq.~\eqref{eq:error-configuration} is thus generated by the $X$-type stabilizers, reducing the gauge parameter $\boldsymbol{\alpha}$ to $r^X$ binary components. Introducing a gauge spin $\sigma_j \in \{\pm 1\}$ on each of the $r^X$ independent $X$-type stabilizer sites maps the per-class likelihood onto a random multi-spin Ising model:
\begin{equation}
    H(\boldsymbol{\sigma}) = -J \sum_{q=1}^{n} \eta_q \prod_{j=1}^{r^X} \sigma_j^{S^X_{j,q}}, \quad \beta J = \tfrac{1}{2}\ln\frac{1-p}{p},
    \label{eq:rbim_general}
\end{equation}
where $\eta_q \equiv (-1)^{(\boldsymbol{\gamma}D + \ell L)^X_q}$ is the quenched disorder, determined by the pure error and the logical class label $\ell$ (ensuring distinct classes correspond to distinct disorder realizations), $S^X_{j,q}$ encodes the support of the $X$-type stabilizer $j$ on qubit $q$, and $\beta$ is the inverse Nishimori temperature~\cite{Nishimori1981}. 

For a fixed logical class $\ell$, summing the Boltzmann factor $e^{-\beta H(\boldsymbol{\sigma})}$ over all gauge-spin configurations $\boldsymbol{\sigma}$ yields the class partition function $Z_\ell = \sum_{\boldsymbol{\sigma}} e^{-\beta H(\boldsymbol{\sigma})}$, so the class likelihood in Eq.~\eqref{eq:MLD} is proportional to $Z_\ell$. Defining the free energy $F_\ell = -\beta^{-1}\ln Z_\ell$, the MLD objective reduces to $\hat{\ell}_{\mathrm{MLD}} = \argmin_\ell F_\ell$, while MWD corresponds to replacing $F_\ell$ by the minimum weight $w_{\ell,1}$. The proposed AMLD framework in Sec.~\ref{sec:algorithm} interpolates between these two limits by constructing a truncated estimate $\hat{F}_\ell$ from a finite candidate pool.

We define the many-body interaction order $R$ as the maximum number of gauge spins appearing in any Hamiltonian term associated with a single physical qubit. 
For the toric code~\cite{Kitaev2003,Dennis2002}, each qubit is incident to two $X$-type stabilizers (vertex operators) and therefore $R=2$. For the $6.6.6$ color code~\cite{Bombin2006,Ohzeki2009}, each bulk qubit is incident to three $X$-type stabilizers, giving $R=3$, corresponding to a random three-body Ising model~\cite{Katzgraber2009}.

Under depolarizing noise, the need to track both $X$- and $Z$-type syndromes requires two gauge sectors $\boldsymbol{\sigma}$ and $\boldsymbol{\mu}$, with the Nishimori condition modified to $\beta J = \tfrac{1}{4}\ln[3(1-p)/p]$~\cite{Bombin2012}. Each $Y$ error triggers both $X$- and $Z$-type syndromes simultaneously, generating a multiplicative cross term that couples the two sectors. For the toric code,
\begin{align}
    H_{\mathrm{toric}}^{\mathrm{dep}}(\boldsymbol{\sigma}, \boldsymbol{\mu}) = -J \sum_{q=1}^{n} \Big[
    &\eta^X_q\, \sigma_{i_q} \sigma_{j_q} + \eta^Z_q\, \mu_{u_q} \mu_{v_q} \notag \\
    + \;&\eta^X_q \eta^Z_q\, \sigma_{i_q} \sigma_{j_q} \mu_{u_q} \mu_{v_q} \Big],
    \label{eq:depol_toric}
\end{align}
where $(i_q, j_q)$ and $(u_q, v_q)$ index the two $X$-type vertices and the two $Z$-type plaquettes adjacent to qubit $q$, respectively. 
This cross term raises the maximum interaction order to $R=4$ for the toric code (two $Z$-plaquettes plus two $X$-vertices per qubit)~\cite{takeuchi2023depolarizingising} and $R=6$ for the $6.6.6$ color code (three faces per qubit per sector)~\cite{Landahl2011}. An alternative formulation, iterative low-order decoding (ILOD), approximates these cross-sector correlations using alternating low-order $X$- and $Z$-type sub-Hamiltonians~\cite{liu2026iterative}.

\section{The AMLD algorithmic framework}
\label{sec:algorithm}

The exact MLD objective in Eq.~\eqref{eq:MLD} requires summation over the full coset $C_\ell$. AMLD circumvents this without modifying the underlying solver. Treating the solver as a black-box stochastic generator, it recycles the otherwise discarded samples to capture partial information about the coset degeneracy. This degeneracy is the entropic contribution that MWD discards but MLD retains.

\subsection{Empirical free-energy truncation}
\label{sec:truncation}

For a given syndrome $\boldsymbol{\gamma}$, the stochastic inner solver is run for $N_{\mathrm{total}}$ trials, producing an empirical ensemble of candidate error configurations. We partition this ensemble into logical equivalence classes and retain only distinct configurations within each class, yielding the per-class pool $S_\ell \subseteq C_\ell$ for each $\ell \in \mathcal{L}$.
Instead of evaluating the intractable sum over the full coset $C_\ell$, AMLD constructs an empirical truncation of the partition function restricted to the discovered subset $S_\ell$:
\begin{equation}
    \hat{Z}_\ell = \sum_{\boldsymbol{e} \in S_\ell} \exp\left( -\beta w(\boldsymbol{e}) \right),
    \label{eq:Z_estimator}
\end{equation}
where $w(\boldsymbol{e})$ is the Pauli weight and $\beta$ is the Nishimori inverse temperature in the weight convention of Sec.~\ref{sec:stabilizer_codes}, defined by $-\ln P(\boldsymbol{e}) = \beta\, w(\boldsymbol{e}) + \mathrm{const}$. Explicitly, $\beta = \ln[(1-p)/p]$ for bit-flip noise and $\beta = \ln[3(1-p)/p]$ for depolarizing noise. AMLD then returns the class minimizing the truncated free energy $\hat{F}_\ell \equiv -\beta^{-1}\ln\hat{Z}_\ell$:
\begin{equation}
    \ell^*_{\mathrm{AMLD}} = \argmin_{\ell \in \mathcal{L}} \hat{F}_\ell.
    \label{eq:DecisionRule}
\end{equation}

The truncated free energy $\hat{F}_\ell$ aggregates the Boltzmann weights of all distinct configurations in $S_\ell$, balancing the empirical minimum weight against the entropic contribution from the multiplicity of discovered low-weight states. AMLD can thereby resolve cross-class minimum-weight ties when the candidate pools reveal different sampled minimum-weight multiplicities; this mechanism is formalized in Sec.~\ref{sec:theory_degeneracy} via a two-class minimal model.

\subsection{Candidate generating strategies}
\label{sec:generation_routes}

Depending on the architecture of the underlying solver and the code structure, the per-class pools $\{S_\ell\}$ can be populated via two strategies:

\textbf{Strategy A: Global Sampling with Class Assignment.}
The solver is run for $N_{\mathrm{total}}$ trials without per-class control, producing a global collection of syndrome-consistent corrections. Each output is then projected to its logical class $\ell$ via the symplectic inner product with the logical observables, and routed to the corresponding pool $S_\ell$. Strategy A is a practical option for codes with large logical-class counts (e.g., the bivariate bicycle codes), where $|\mathcal{L}|$ grows exponentially with the logical-qubit count $k$ and class-by-class sampling becomes prohibitively expensive. We adopt strategy A for our BP-OSD and RW experiments.

\textbf{Strategy B: Class-Restricted Sampling.} 
For topological codes with a small number of logical classes (e.g., $|\mathcal{L}|=4$ for the toric code under bit-flip noise, $|\mathcal{L}|=16$ under depolarizing noise), the sampling can be explicitly decomposed. For each class $\ell$, we fix a logical-operator representative and run the solver constrained to that class for $N$ independent trials. This yields $N_{\mathrm{total}} = N|\mathcal{L}|$ total runs. While incurring a computational overhead linear in $|\mathcal{L}|$, this route ensures that every logical class receives equal sampling effort. We adopt this strategy for the Ising-decoder experiments based on the Hamiltonians defined in Sec.~\ref{sec:mapping}.

\subsection{Algorithm and complexity}
\label{sec:algo_complexity}

Algorithm~\ref{alg:amld} describes the decoding procedure. After the candidates are generated, the post-processing reduces to deduplication within each pool followed by a log-sum-exp evaluation. For a finite nonempty set $\{x_i\}$, we evaluate the log-sum-exp using the numerically stable identity
\begin{equation*}
\begin{aligned}
m &\equiv \max_i x_i,\\
\mathrm{logsumexp}\!\left(\{x_i\}\right)
&\equiv m+\ln\sum_i \exp(x_i-m).
\end{aligned}
\end{equation*}

\begin{algorithm}[htbp]
\linespread{1.2}\selectfont

\caption{Approximate maximum-likelihood decoding (AMLD)\label{alg:amld}}
\SetKwInOut{KwParam}{Parameters}
\KwParam{channel error rate $p$; stochastic inner solver; trial budget $N_{\mathrm{total}}$}
\KwIn{syndrome $\boldsymbol{\gamma}$}
\KwOut{Recovery operation $\boldsymbol{e}_{\mathrm{AMLD}}$}
$\beta \leftarrow $ Nishimori inverse temperature for error rate $p$\;

Run the inner solver for $N_{\mathrm{total}}$ trials, collect the generated candidate configurations, filter for unique instances, and aggregate them into per-class pools $\{S_\ell\}_{\ell \in \mathcal{L}}$\;

Initialize $\hat{F}_\ell \leftarrow +\infty$ for every $\ell \in \mathcal{L}$\;
\ForEach{non-empty class $\ell \in \mathcal{L}$}{ 
    $\ln\hat{Z}_\ell \leftarrow \mathrm{logsumexp}\left(\{-\beta w(\boldsymbol{e}) : \boldsymbol{e} \in S_\ell\}\right)$ \tcp*{numerically stable summation}
    $\hat{F}_\ell \leftarrow -\beta^{-1} \ln\hat{Z}_\ell$\;
}
$\ell^* \leftarrow \argmin_{\ell} \hat{F}_\ell$\;
\Return{$\argmin_{\boldsymbol{e} \in S_{\ell^*}} w(\boldsymbol{e})$ }\;
\end{algorithm}

Let $N_{\mathrm{cand}}$ denote the total number of candidate configurations generated before deduplication. The AMLD-specific post-processing consists of logical-class projection ($\mathcal{O}(knN_{\mathrm{cand}})$ for Strategy~A), unique-element hashing with expected cost $\mathcal{O}(nN_{\mathrm{cand}})$, and log-sum-exp evaluations with total cost $\mathcal{O}(N_{\mathrm{cand}})$. The overall post-processing cost is therefore $\mathcal{O}((kn+n)N_{\mathrm{cand}})$. For the code instances studied here, $k$ is fixed and small, so this reduces to $\mathcal{O}(nN_{\mathrm{cand}})$.

\section{Theoretical analysis}
\label{sec:theory}

We first derive a two-sided bound on $\hat{F}_\ell$ that interpolates between MWD and MLD and then analyze the degeneracy-resolution mechanism in a minimal two-class setting. Finally, we examine finite-sampling convergence and clarify that, in the ground-state-dominated regime, AMLD can use nonthermal samples provided that the sampler discovers sufficiently diverse low-weight configurations.

\subsection{Truncation-error bounds and MLD recovery}
\label{sec:theory_bounds}

AMLD approximates the exact partition function $Z_\ell$ via an empirical subset $S_\ell \subseteq C_\ell$. For any non-empty pool ($|S_\ell| \geq 1$) and $\beta>0$, the truncated free-energy estimator $\hat{F}_\ell = -\beta^{-1}\ln\hat{Z}_\ell$ satisfies a two-sided bound:
\begin{equation}
    F_\ell \leq \hat{F}_\ell \leq \hat{w}_{\ell,1},
    \label{eq:two_sided_bound}
\end{equation}
where $F_\ell$ is the exact free energy and $\hat{w}_{\ell,1} \equiv \min_{\boldsymbol{e} \in S_\ell} w(\boldsymbol{e})$ is the empirical minimum weight in $S_\ell$.

\textit{Proof.}---The lower bound follows from the positivity of Boltzmann factors. Since $S_\ell \subseteq C_\ell$, the truncated partition function satisfies $\hat{Z}_\ell \leq Z_\ell$, which implies $\hat{F}_\ell \geq F_\ell$. The lower bound saturates for class $\ell$ when $S_\ell=C_\ell$; AMLD recovers exact MLD when this holds for every logical class. For the upper bound, the configuration achieving the empirical minimum weight in $S_\ell$ contributes a single term $\exp(-\beta \hat{w}_{\ell,1})$ to the sum $\hat{Z}_\ell$, yielding $\hat{Z}_\ell \geq \exp(-\beta \hat{w}_{\ell,1})$ and thus $\hat{F}_\ell \leq \hat{w}_{\ell,1}$. The upper bound saturates for class $\ell$ when $|S_\ell|=1$; AMLD reduces to a minimum-weight decision over the sampled candidates when this holds for every represented logical class.\hfill$\square$

Because AMLD decides via $\argmin_\ell \hat{F}_\ell$, recovering the exact MLD output does not require full reconstruction of $Z_\ell$. Assuming a unique MLD-optimal class $\ell^* \equiv \argmin_\ell F_\ell$, it suffices that its truncation error remains smaller than the inter-class free-energy gap $\Delta F \equiv \min_{\ell \neq \ell^*} (F_\ell - F_{\ell^*})$:
\begin{equation}
    0 \leq \hat{F}_{\ell^*} - F_{\ell^*} < \Delta F.
    \label{eq:sufficient}
\end{equation}
Since Eq.~\eqref{eq:two_sided_bound} guarantees $\hat{F}_\ell \geq F_\ell$ for all classes, Eq.~\eqref{eq:sufficient} is a sufficient condition to ensure $\hat{F}_{\ell^*} < \hat{F}_\ell$ for every $\ell \neq \ell^*$, enabling AMLD to recover the MLD decision for this syndrome.

\subsection{Degeneracy resolution: A two-class analysis}
\label{sec:theory_degeneracy}

The advantage of AMLD over MWD is most pronounced in the degenerate regime. Consider two logical classes, $\ell_1$ and $\ell_\chi$, sharing the same global minimum weight $w_{\min}$. Let $\ell_1$ contain a unique minimum-weight configuration, while $\ell_\chi$ has a ground-state degeneracy $\chi \geq 2$, \emph{i.e.}, $\chi$ distinct minimum-weight configurations. For both classes, assume the first-excited level lies at $w_{\min} + \Delta$.

MWD only selects the minimum weight and is insensitive to the ground-state degeneracy $\chi$, yielding $P_{\mathrm{MWD}} = 1/2$ under random tie-breaking. In the low-noise limit, MLD accounts for the ground-state degeneracy and favors $\ell_\chi$.
To analyze AMLD's resolution of this tie, we assume a class-restricted sampling strategy (Strategy B in Sec.~\ref{sec:generation_routes}) drawing $N$ independent trials per class. For a thermal solver operating at the Nishimori temperature parameterized by the channel error rate $p$, the probability of drawing the unique ground state in $\ell_1$ is $P_1 = [1 + \mathcal{O}(p^\Delta)]^{-1}$, where the implicit constant is proportional to the first-excited state degeneracy. For the degenerate class $\ell_\chi$, the $\chi$ equivalent ground states collectively enhance the single-trial capture probability to $P_\chi = [1 + \mathcal{O}(p^\Delta/\chi)]^{-1}$.

In the low-noise regime, conditioned on each class yielding at least one global-minimum configuration, the truncated free energy is dominated by the number of distinct global-minimum configurations discovered: $\hat{F}_\ell \approx w_{\min} - \beta^{-1}\ln \hat{n}_{\ell,1}$, where $\hat{n}_{\ell,1} \equiv |\{\boldsymbol{e} \in S_\ell : w(\boldsymbol{e}) = w_{\min}\}|$. Because $\ell_1$ contains a unique global-minimum configuration, $\hat{n}_{\ell_1,1}=1$ under this conditioning. Therefore, drawing at least two \emph{distinct} global-minimum configurations from $\ell_\chi$ guarantees $\hat{n}_{\ell_\chi,1} \geq 2 > \hat{n}_{\ell_1,1}$, ensuring that AMLD selects $\ell_\chi$ without tie-breaking.
The probability of this strict-resolution outcome within $N$ trials is:
\begin{equation}
    P_N^{(\mathrm{strict})} = 1 - \chi \left( 1 - P_\chi + \frac{P_\chi}{\chi} \right)^N + (\chi-1)(1-P_\chi)^N.
    \label{eq:PN_exact}
\end{equation}
For the minimal budget $N=2$, Eq.~\eqref{eq:PN_exact} gives $P_2^{(\mathrm{strict})} \to 1-1/\chi$ as $P_\chi\to1$. In the joint low-noise limit $P_1,P_\chi\to1$, the probabilities that either class yields no global-minimum configuration vanish. Failure of strict resolution is then dominated by the event $\hat{n}_{\ell_1,1}=\hat{n}_{\ell_\chi,1}=1$, for which AMLD uses the same random tie-breaking as MWD. Thus, as $p\to0$, $P_N^{\mathrm{AMLD}}=\tfrac{1}{2}(1+P_N^{(\mathrm{strict})})+o(1)$. The leading-order success probability exceeds the MWD value of $1/2$ for $N\geq2$ and $\chi\geq2$ and increases monotonically with $N$ and $\chi$.

\subsection{Finite-sampling convergence and robustness}
\label{sec:theory_robustness}

\begin{figure}[htbp]
    \centering
    \includegraphics[width=0.66\linewidth]{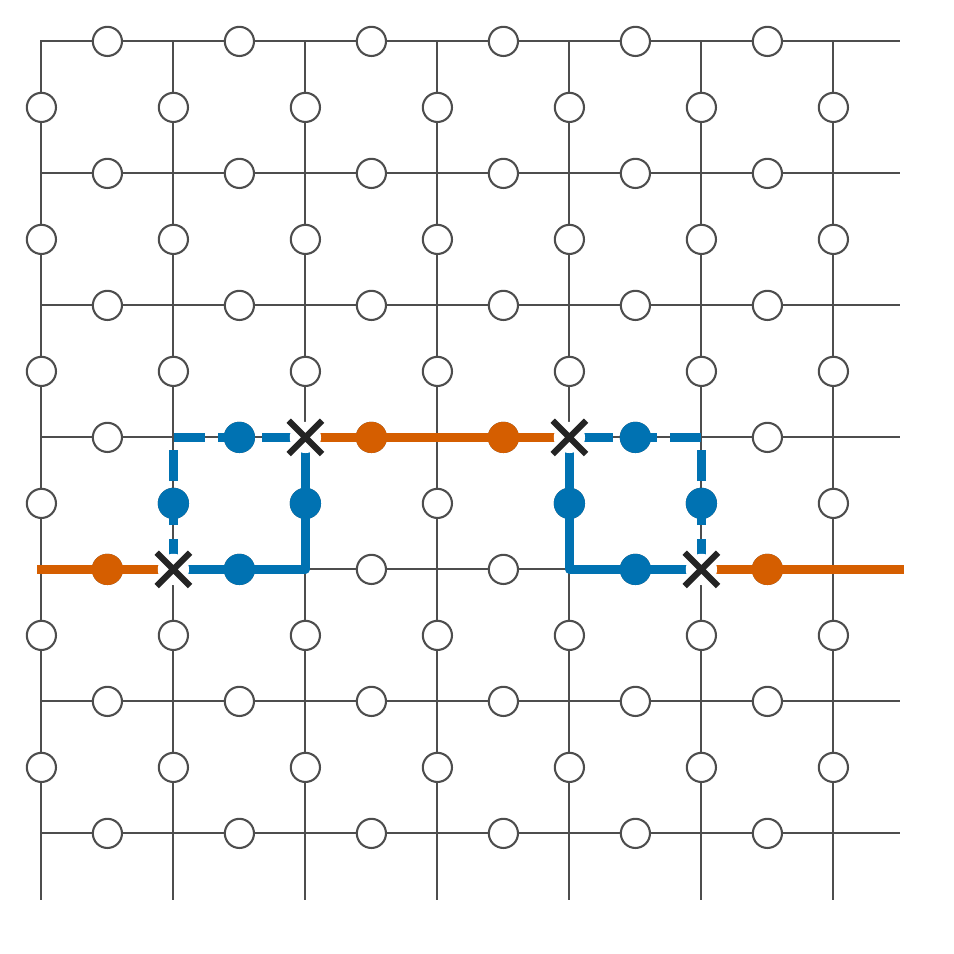}
    \caption{Two logical classes associated with a degenerate syndrome on the $6\times 6$ toric code. This syndrome is compatible with minimum-weight error chains belonging to two distinct logical classes. The blue class $\ell_\chi$ has a ground-state degeneracy of $\chi=4$ minimum-weight configurations (generated from the displayed chain by the independent face stabilizers marked with dashed blue edges), while the red class $\ell_1$ has a unique minimum-weight configuration.} 
    \label{fig:advantage_example}
\end{figure}

Within the two-class low-noise model, Eq.~\eqref{eq:PN_exact} shows that full coset enumeration is unnecessary for strict tie resolution: discovering at least two distinct ground states in $\ell_\chi$ is sufficient for AMLD to select the MLD-preferred class. Figure~\ref{fig:advantage_example} shows an example on the $6\times 6$ toric code that realizes this two-class degeneracy.
We use it to examine the convergence explicitly.

\begin{figure}[htbp]
    \centering
    \includegraphics[width=0.8\linewidth]{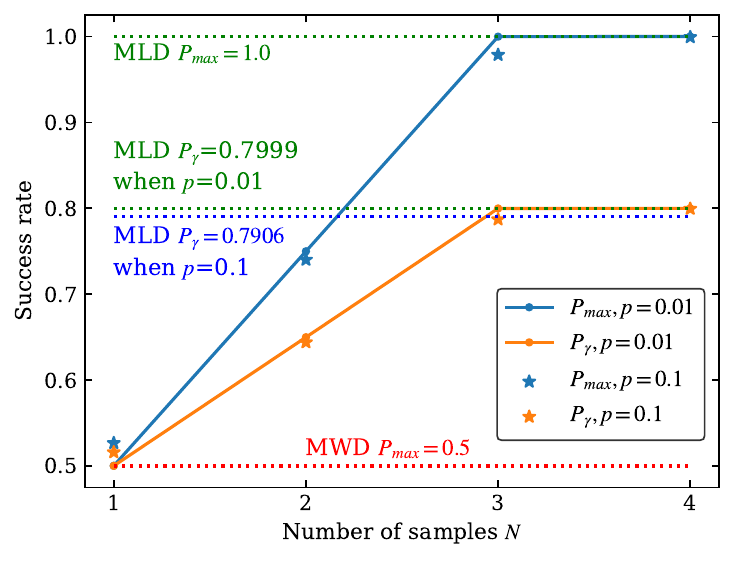}
    \caption{Decoding success probabilities for the syndrome configuration in Fig.~\ref{fig:advantage_example} as a function of trials $N$ per class. The true physical noise is fixed at $p_0 = 0.01$, while the Ising solver operates at two effective noise levels: $p'=0.01$ (exact Nishimori temperature) and $p'=0.1$ (mismatched prior). For both $P_{\max}$ and $P_\gamma$, AMLD converges to the MLD limit by $N=3$. The performance under the mismatched prior $p'=0.1$ (stars) remains marginally below that under the matched prior $p'=0.01$, demonstrating robustness to thermal miscalibration.}
    \label{fig:amld-trials}
\end{figure}

For the syndrome configuration in Fig.~\ref{fig:advantage_example}, explicit enumeration of the low-weight errors yields the following weight spectra for the two competing classes (the first excited level lies at $w_{\min}+\Delta$ with $\Delta=2$ for both classes):
\begin{align}
    \mathbf{w}_{\ell_1}  &= \{4,6,6,6,6,6,6,6,6,8,8,8,8,\dots\}, \notag \\
    \mathbf{w}_{\ell_\chi} &= \{4,4,4,4,6,6,6,6,6,6,6,6,8,\dots\}.
    \label{eq:weight_lists}
\end{align}
The ellipses omit higher-weight contributions, whose Boltzmann factors are exponentially suppressed and contribute negligibly to $\hat{Z}_\ell$ in the parameter regime of Fig.~\ref{fig:amld-trials}.

Evaluating the truncated estimator on these exact weight spectra yields the success probabilities shown in Fig.~\ref{fig:amld-trials}. We track two metrics: $P_{\max}$, the probability that AMLD selects the class of largest exact posterior probability, and $P_\gamma$, the probability that AMLD selects the class containing the true error. Both probabilities rise sharply from the MWD baseline ($N=1$) and reach the MLD limit by $N=3$ samples per class. For this example, these results show that $|S_\ell| \ll |C_\ell|$ is sufficient for the truncation error to fall below the inter-class free-energy gap.

To verify that the degeneracy-resolution mechanism persists in the near-threshold regime, we examine the SA dataset on the toric code under bit-flip noise at $p=0.100$ and depolarizing noise at $p=0.170$. For every syndrome where AMLD and MWD select different classes, we record the empirical minimum-weight difference $\Delta w \equiv \hat{w}_{\ell^*_{\mathrm{AMLD}},1} - \hat{w}_{\ell^*_{\mathrm{MWD}},1}$. We observe $\Delta w = 0$ in $88.8\%$ of these cases under bit-flip noise and $99.6\%$ under depolarizing noise. The dominant decision mechanism is therefore the resolution of cross-class minimum-weight ties, rather than entropy-driven selection of classes with strictly higher empirical minimum weight. The residual $11.2\%$ with $\Delta w > 0$ under bit-flip noise reflects the finite-temperature entropy--energy competition intrinsic to $\hat{F}_\ell$: at $\beta < \infty$, the truncated free energy can favor a class with a strictly higher empirical minimum weight when its aggregated entropic contribution (from many low-weight states) overcomes the energy penalty. 

In the ground-state-dominated regime, the truncated partition function of any class sharing the global minimum weight scales as $\hat{Z}_\ell \sim \hat{n}_{\ell,1}\exp(-\beta w_{\min})$, giving $\hat{F}_\ell \approx w_{\min}-\beta^{-1}\ln\hat{n}_{\ell,1}$. For a fixed candidate pool, the shared $w_{\min}$ cancels in the cross-class comparison, reducing the decision to $\argmax_\ell\hat{n}_{\ell,1}$; this ranking is independent of the positive value of $\beta$ used to evaluate the pool.

This fixed-pool independence does not imply that candidate generation is independent of the solver parameter $\beta'$, because changing $\beta'$ can change which minimum-weight configurations are discovered and hence the sampled multiplicities $\hat{n}_{\ell,1}$. Figure~\ref{fig:amld-trials} nevertheless provides empirical evidence of robustness to the tested mismatch: the success rates at $p'=0.1$ remain close to those at the matched value $p'=0.01$. More broadly, AMLD does not require thermally distributed samples, provided that the underlying sampler discovers sufficiently diverse low-weight configurations. This observation motivates evaluating AMLD with RW and BP-OSD in Sec.~\ref{sec:numerics}, as neither method relies on thermal sampling to generate candidates.

\section{Numerical results}
\label{sec:numerics}

We evaluate the AMLD framework using three families of decoding solvers: an SA-based Ising decoder for the toric and color codes under bit-flip and depolarizing noise, RW for the toric code under bit-flip noise, and BP-OSD for the BB code. Throughout this section, AMLD and the MWD baseline operate on identical sample pools. SA on topological codes with small $|\mathcal{L}|$ uses Strategy~B: $N$ trials per class, $N_{\mathrm{total}} = N|\mathcal{L}|$. RW on the toric code and BP-OSD on the BB code use Strategy~A: the candidate outputs of $N_{\mathrm{total}}$ global trials are directly classified into per-class pools. In both cases, MWD retains the single lowest-weight configuration across all runs. Because AMLD and MWD operate on the same candidate pools, their comparison isolates the effect of the final decision rule. Appendix~\ref{sm:solvers} details solver-specific implementations; Appendix~\ref{sm:threshold} describes the threshold-extraction procedure.

\subsection{Threshold improvements}
\label{sec:sa_threshold}

\begin{figure*}[htbp]
    \centering
    \begin{minipage}[t]{0.42\linewidth}
        \centering
        \includegraphics[width=\linewidth]{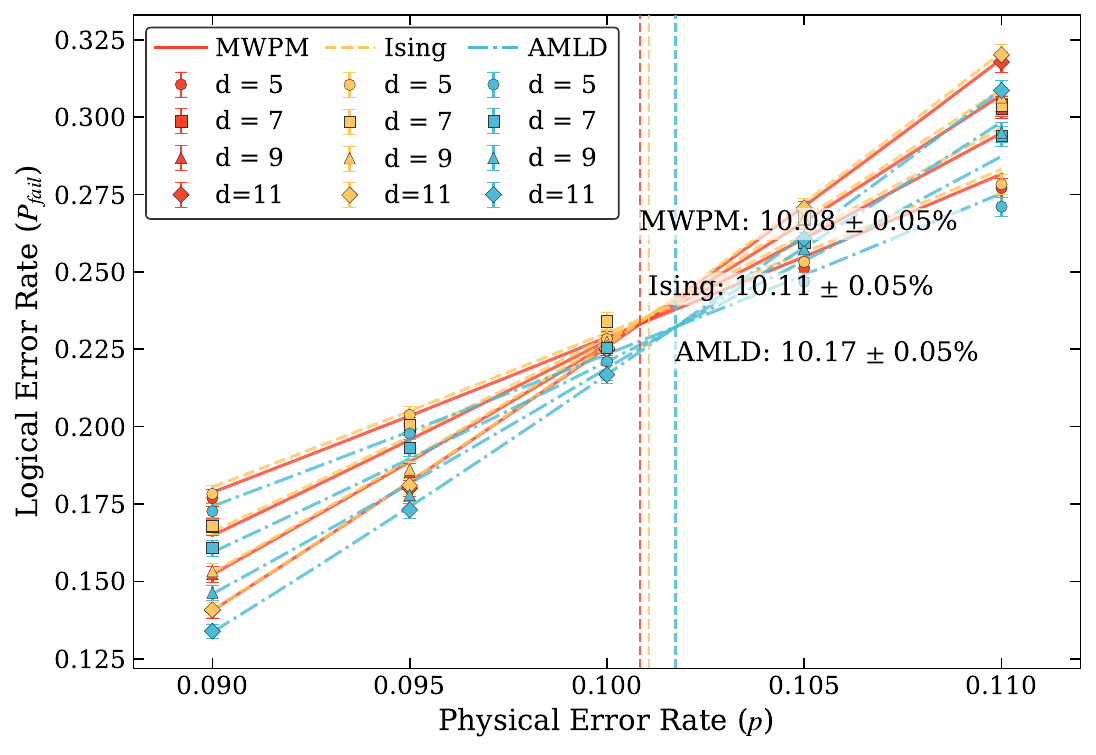}
        \par(a) Toric code (bit-flip noise)
    \end{minipage}
    \hfill
    \begin{minipage}[t]{0.42\linewidth}
        \centering
        \includegraphics[width=\linewidth]{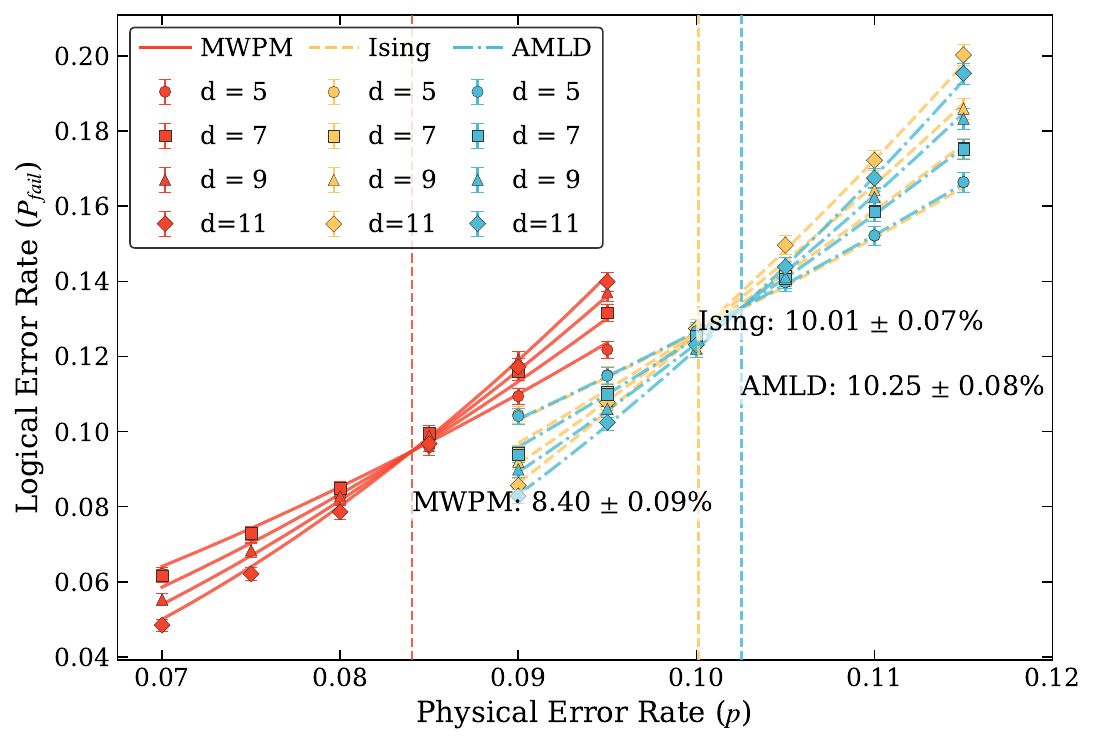}
        \par(b) $6.6.6$ color code (bit-flip noise)
    \end{minipage}
    \begin{minipage}[t]{0.42\linewidth}
        \centering
        \includegraphics[width=\linewidth]{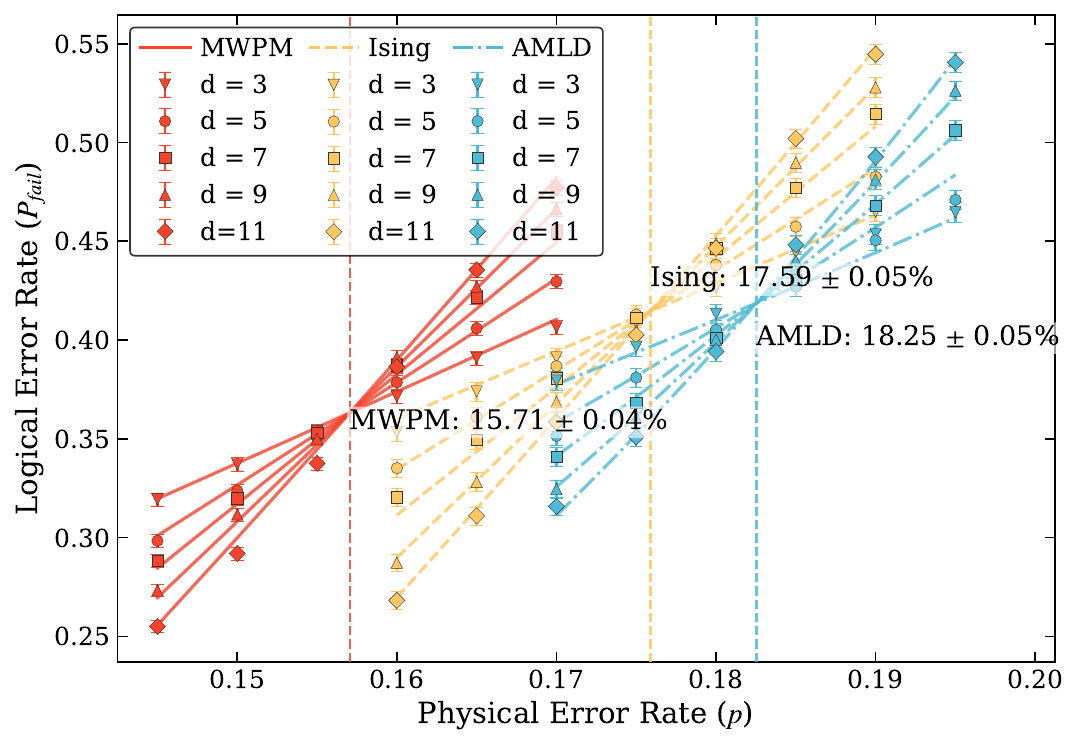}
        \par(c) Toric code (depolarizing noise)
    \end{minipage}
    \hfill
    \begin{minipage}[t]{0.42\linewidth}
        \centering
        \includegraphics[width=\linewidth]{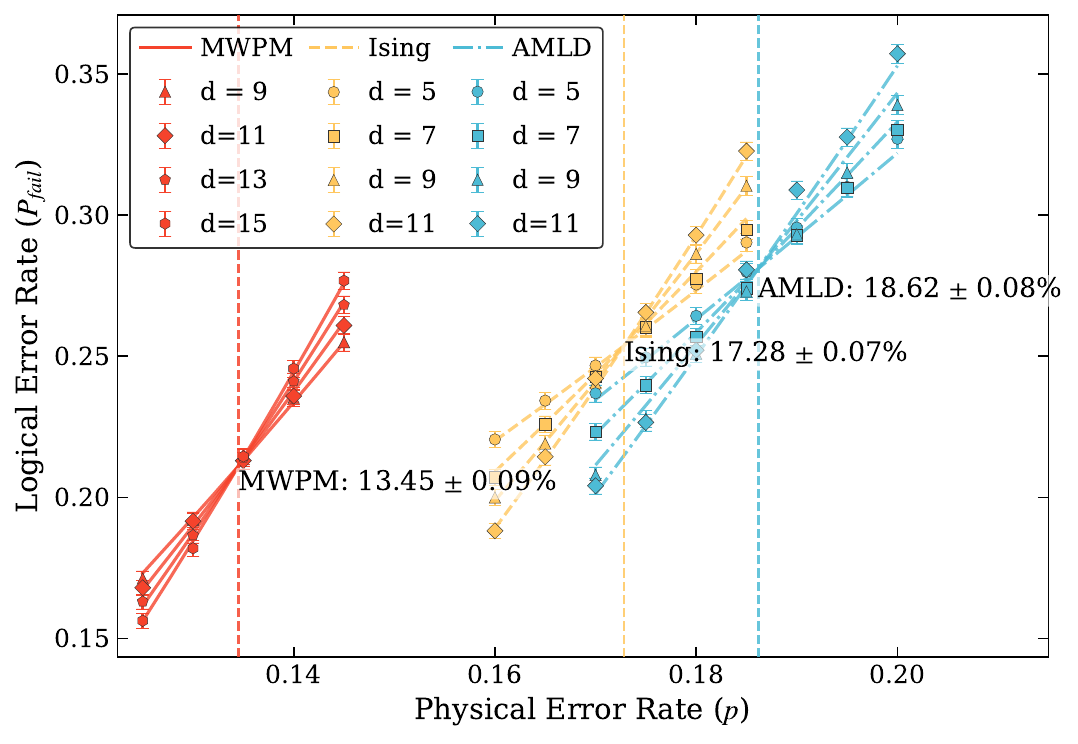}
        \par(d) $6.6.6$ color code (depolarizing noise)
    \end{minipage}
    \caption{Logical error rate versus physical error rate for Ising decoding based on an SA solver, comparing AMLD[$N$], MWD[$N$], and MWPM, with $N=20$ runs per logical class. Threshold crossings are indicated by the vertical bands.}
    \label{fig:thresholds_all}
\end{figure*}

Applying SA to the Ising Hamiltonians defined in Sec.~\ref{sec:mapping}, we obtain the fitted threshold estimates shown in Fig.~\ref{fig:thresholds_all} with the standard finite-size-scaling ansatz. The thresholds of the minimum-weight perfect matching (MWPM) decoder with and without considering correlations for the toric code~\cite{higgott2025sparse} and the color code~\cite{sahay2022colormwpm} are also plotted for comparison.

The AMLD/MWD thresholds are summarized in Table~\ref{tab:sa_results}.
The largest improvement occurs on the $6.6.6$ color code under depolarizing noise, where AMLD raises the threshold from $17.28(7)\%$ to $18.62(8)\%$, closing $83\%$ of the gap to the theoretical $18.9\%$ MLD limit~\cite{Bombin2012,BravyiSucharaVargo2014}.
For the toric code under depolarizing noise, the threshold rises from $17.59(5)\%$ to $18.25(5)\%$, closing $50\%$ of the corresponding gap. 
Under bit-flip noise, AMLD lifts the threshold of the toric code from $10.11(5)\%$ to $10.17(5)\%$ and the threshold of the color code from $10.01(7)\%$ to $10.25(8)\%$, closing $7\%$ and $27\%$ of the gaps to the Nishimori-line thresholds at $10.93(2)\%$~\cite{Honecker2001,Dennis2002} and $10.9(2)\%$~\cite{Katzgraber2009}, respectively.
The toric-code shift lies within the combined fit uncertainty and is not statistically significant on its own.

\begin{table}[htbp]
\caption{SA + AMLD across four (code, noise) combinations, ordered by $R$. Closure: fraction of the MWD--MLD threshold gap closed by AMLD; $\lambda \equiv p_L(\mathrm{AMLD})/p_L(\mathrm{MWD})$ at $p=0.10$, $d=7$, evaluated at $N=20$ SA runs per logical class as in Fig.~\ref{fig:gain_L}.}
\label{tab:sa_results}
\renewcommand{\arraystretch}{1.2}
\setlength{\tabcolsep}{3pt}  
\begin{tabular}{llcccccc}
\hline\hline
Code & Noise & $R$ & $|\mathcal{L}|$ & $p_{\mathrm{th}}^{\mathrm{MWD}}$ & $p_{\mathrm{th}}^{\mathrm{AMLD}}$ & Closure & $\lambda$ \\
\hline
toric & bit-flip & 2 & 4  & $10.11(5)\%$ & $10.17(5)\%$ & $7\%$  & 0.969 \\
color & bit-flip & 3 & 2  & $10.01(7)\%$ & $10.25(8)\%$ & $27\%$ & 1.00  \\
toric & depol.   & 4 & 16 & $17.59(5)\%$ & $18.25(5)\%$ & $50\%$ & 0.744 \\
color & depol.   & 6 & 4  & $17.28(7)\%$ & $18.62(8)\%$ & $83\%$ & 0.880 \\
\hline\hline
\end{tabular}
\end{table}

\subsection{Empirical trends with $R$ and $|\mathcal{L}|$}
\label{sec:scaling_analysis}

Across the four code--noise settings, $R$ and $|\mathcal{L}|$ are not varied independently; we therefore report empirical trends rather than isolated effects of either quantity.

\begin{figure}[htb]
    \centering
    \includegraphics[width=0.82\linewidth]{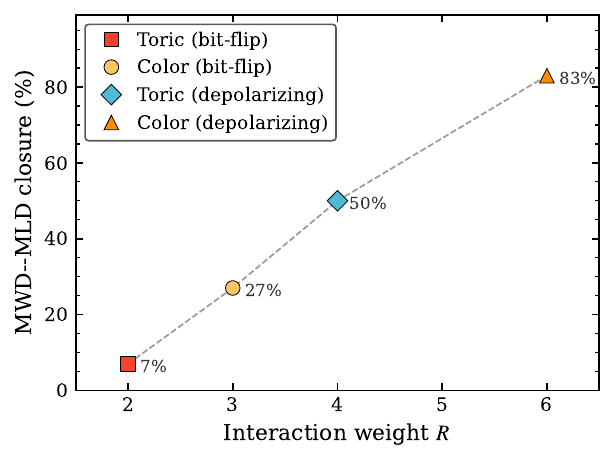}
    \caption{Fraction of the MWD--MLD threshold gap closed by AMLD versus the maximum interaction order $R$ for the four code--noise settings considered.} 
    \label{fig:closure_R}
\end{figure}

\textit{Interaction order $R$.} The fraction of the MWD--MLD threshold gap closed by AMLD increases monotonically with $R$ across the code--noise settings shown in Fig.~\ref{fig:closure_R}, from $7\%$ at $R=2$ (toric, bit-flip) to $83\%$ at $R=6$ (color, depolarizing). 
Within each code, both $R$ and this fraction increase when moving from bit-flip to depolarizing noise: from $7\%$ at $R=2$ to $50\%$ at $R=4$ for the toric code, and from $27\%$ at $R=3$ to $83\%$ at $R=6$ for the color code.
This trend is consistent with the degeneracy-resolution mechanism (Sec.~\ref{sec:theory_degeneracy}): in the ground-state-dominated regime, AMLD can benefit when cross-class minimum-weight ties occur for a larger fraction of syndrome instances and when the tied classes have different sampled ground-state multiplicities $\hat{n}_{\ell,1}$.

\begin{figure}[htb]
    \centering
    \includegraphics[width=0.82\linewidth]{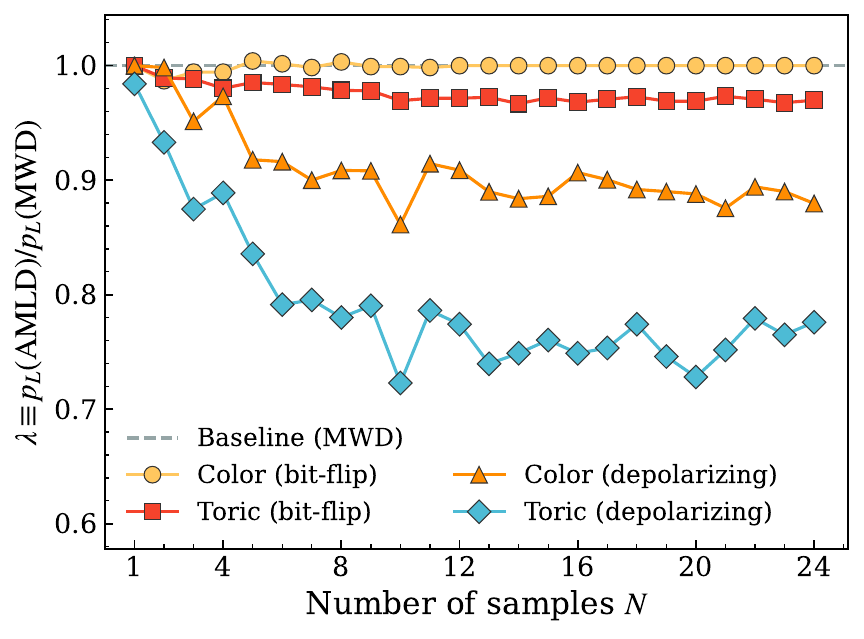}
    \caption{Logical-error-rate ratio $\lambda \equiv p_L(\mathrm{AMLD})/p_L(\mathrm{MWD})$ versus the number $N$ of SA runs per logical class at $p=0.10$ and $d=7$, grouped by $|\mathcal{L}|$. The dashed line marks the MWD baseline ($\lambda=1$); values below one indicate a lower logical error rate for AMLD.}
    \label{fig:gain_L}
\end{figure}

\textit{Logical-class count $|\mathcal{L}|$.}---We evaluate the logical-error-rate ratio $\lambda \equiv p_L(\mathrm{AMLD})/p_L(\mathrm{MWD})$ at $p=0.10$ and $d=7$; the values reported below correspond to $N=20$ SA runs per logical class. This is a logical-error-rate metric rather than a threshold metric: $\lambda<1$ indicates a lower logical error rate for AMLD. Because $p=0.10$ lies near the thresholds of the bit-flip models but well below those of the depolarizing models, we compare the color and toric codes only within the same noise model. Under bit-flip noise, $\lambda\approx1.00$ for the color code ($|\mathcal{L}|=2$) and $\lambda=0.969$ for the toric code ($|\mathcal{L}|=4$). Under depolarizing noise, $\lambda=0.880$ for the color code ($|\mathcal{L}|=4$) and $\lambda=0.744$ for the toric code ($|\mathcal{L}|=16$) (Fig.~\ref{fig:gain_L}). Within each noise model, the setting with the larger $|\mathcal{L}|$ has the smaller $\lambda$. The $|\mathcal{L}|=2$ case shows no measurable improvement ($\lambda=1.00$ within uncertainty). The BP-OSD benchmark at $|\mathcal{L}|=4096$ is reported separately in Sec.~\ref{sec:generality} because it uses a different code, sampling protocol, and operating point and therefore cannot be compared directly with these values of $\lambda$.

\subsection{Decoder generality}
\label{sec:generality}

The analysis in Sec.~\ref{sec:theory_robustness} shows that AMLD does not require thermally distributed candidates; in the ground-state-dominated regime, it instead requires the sampler to discover sufficiently diverse low-weight configurations. This motivates evaluating AMLD with decoders whose output distributions deviate from the Boltzmann form.

\begin{figure}[htb]
    \centering
    \includegraphics[width=0.82\linewidth]{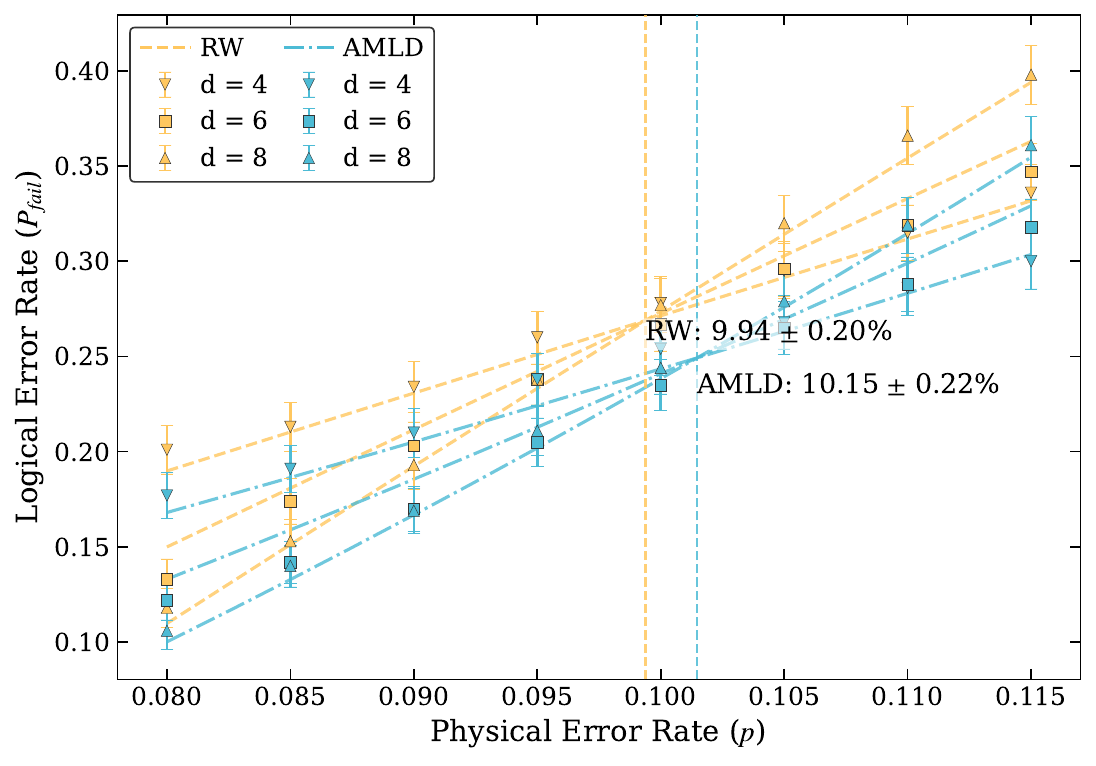}
    \caption{Logical error rate versus physical error rate for RW-based decoding of the toric code under bit-flip noise, comparing AMLD and MWD.}
    \label{fig:rw_threshold}
\end{figure}

\textit{Random Windows (RW).}---RW~\cite{dumer2017distance,pryadko2022qdistrnd} is a probabilistic linear-system decoder whose stochasticity originates from randomized support for the decoded codeword, mainly induced by random column permutations of the parity-check matrix. The decoder's effectiveness has been demonstrated in many recent work on discovering new qLDPC codes with high distances
\cite{scruby2026HighThresholdLowOverheadSingleShot,yang2026PlanarFaulttolerantLogical,webster2026DistanceFindingAlgorithmsQuantuma,bhardwaj2026HighrateQLDPCProcessors,gu2026NearestneighbourGatesArea,tripier2026FaultTolerantQuantumComputinga}.
On the toric code under bit-flip noise, applying AMLD to RW yields a threshold estimate of $10.15\%$, statistically indistinguishable from the MWD estimate of $9.94\%$ for code distances $d=4,6,8$ (Fig.~\ref{fig:rw_threshold}). Thus, this benchmark shows neither a measurable AMLD advantage nor a statistically significant degradation relative to the shared-pool MWD baseline.

\textit{BP-OSD on the BB code.}---Unlike RW, BP-OSD~\cite{Roffe2020} is near-deterministic, so repeated decoding trials require externally induced variation to generate a diverse candidate pool. We induce trial-dependent variation in the decoder prior by multiplying the decoder-assumed per-qubit error probabilities by independent lognormal factors ($\sigma=0.25$), while keeping the physical error rate unchanged and the min-sum scaling factor fixed at $0.625$.
In our toric-code tests ($|\mathcal{L}|=4$), modest prior perturbations typically returned the same correction and therefore generated insufficient within-class candidate diversity for AMLD to differ measurably from MWD. We consequently evaluate perturbed BP-OSD on the BB code ($|\mathcal{L}|=4096$), whose hypergraph-structured decoding problem admits multiple competing low-weight corrections. There, the perturbed prior steers BP-OSD toward different corrections across runs, populating the per-class pools with diverse candidates and enabling AMLD to evaluate the truncated free energies.

\begin{figure}[htb]
    \centering
    \includegraphics[width=0.82\linewidth]{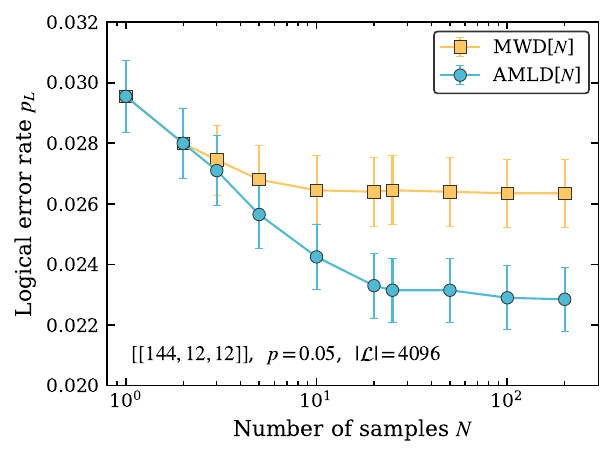}
    \caption{Logical error rate versus trial count $N_{\mathrm{total}}$ for BP-OSD on the $[[144, 12, 12]]$ BB code ($|\mathcal{L}| = 4096$) at $p = 0.05$. MWD saturates by $N_{\mathrm{total}} \approx 20$, while AMLD continues to improve with additional trials, achieving a $13\%$ reduction in logical error rate at $N_{\mathrm{total}}=25$.}
    \label{fig:bposd_convergence}
\end{figure}

Figure~\ref{fig:bposd_convergence} shows the logical error rate at $p=0.05$ as a function of the global trial count $N_{\mathrm{total}}$. The performance of MWD saturates at $p_L = 0.0264$ for $N_{\mathrm{total}} \geq 20$. AMLD instead continues to refine the truncated per-class partition-function estimates, reaching $p_L = 0.0230$ at $N_{\mathrm{total}} = 25$ ($13\%$ reduction relative to MWD) and remaining stable thereafter. These results illustrate how AMLD exploits additional trials: even when their candidate outputs do not improve the minimum weight found by MWD, they can still contribute entropic information to the truncated free-energy estimate.

\section{Discussion and conclusion}
\label{sec:discussion}

This work introduces AMLD, a black-box post-processing framework that reuses candidate configurations from stochastic minimum-weight decoders to construct truncated per-class partition functions. For each nonempty candidate pool, the estimated free energy is bounded below by the exact free energy and above by the empirical minimum weight. It reduces to a minimum-weight decision over the sampled candidates when one candidate is retained per class and recovers exact MLD when the candidate pools cover the full cosets. In the ground-state-dominated regime, AMLD can use differences in sampled ground-state multiplicities to resolve cross-class minimum-weight ties that MWD cannot distinguish.

Benchmarks with SA, RW, and perturbed BP-OSD demonstrate that the same post-processing rule can be applied to Ising, linear-system, and message-passing candidate generators without requiring thermal sampling or matching-based enumeration. The experiments span topological codes and a bivariate-bicycle quantum LDPC code, indicating that candidate pools containing multiple low-weight configurations can provide useful information for AMLD's truncated free-energy estimates.

The numerical results suggest that larger $|\mathcal{L}|$ and $R$ are associated with a greater AMLD advantage, although neither quantity is varied independently of the code family and noise model. Here, $|\mathcal{L}|$ grows exponentially with the logical-qubit count $k$, while, for the mappings studied, $R$ increases when moving from bit-flip to depolarizing noise. The per-class bounds of Eq.~\eqref{eq:two_sided_bound} hold for arbitrary $|\mathcal{L}|$; extending the two-class strict-resolution probability in Eq.~\eqref{eq:PN_exact} to general $|\mathcal{L}|$ remains open.

For circuit-level correlated noise, BP-OSD and RW pipelines could retain the same AMLD post-processing form, with $\beta w(\boldsymbol{e})$ replaced by a channel-consistent negative log-likelihood $-\ln P(\boldsymbol{e})$ evaluated on the spatiotemporal Tanner graph~\cite{gidney2021stim,higgott2025sparse}. Extending the present SA implementation would require an Ising mapping that accommodates circuit-level disorder; developing this extension remains future work. The present results show that samples discarded by minimum-weight selection can still provide useful information for approximate maximum-likelihood decoding.

\begin{acknowledgments}
We acknowledge the support from QUANTA (\textbf{QU}antum f\textbf{AN}s from I\textbf{T} \textbf{A}rea) group. This work has been supported by the National Key R\&D Program of China (Grant No. 2024YFB4504001), the National Natural Science Foundation of China (Grant No. 62302395, 62301505, and 62421002), the Fundamental and Interdisciplinary Disciplines Breakthrough Plan of the Ministry of Education of China (Grant No. JYB2025XDXM202), the Aid Program for Science and Technology Innovative Research Team in Higher Educational Institutions of Hunan Province, and the Innovation Research Foundation of NUDT (Grant No. 202401-YJRC-ZZ-005) \looseness=-1
\end{acknowledgments}

\appendix

\section{Solver implementations}
\label{sm:solvers}

\subsection{Ising decoder by simulated annealing}
\label{sm:sa}

The per-class Ising Hamiltonians (Sec.~\ref{sec:mapping}) are solved by SA using the OpenJij library~\cite{openjij} with the heat-bath updater. The cooling schedule is exponential between fixed $\beta_{\mathrm{anneal}}^{\min} = 0.01$ and $\beta_{\mathrm{anneal}}^{\max} = 10$, where $\beta_{\mathrm{anneal}}$ is the inverse temperature in the SA-internal Markov-chain process, distinct from the channel-matched Nishimori $\beta$ used in the AMLD post-processing [Eq.~\eqref{eq:Z_estimator}]. The number of sweeps per run is scaled with the code distance $d$ until the logical error rate at each $(d, p)$ point saturates; the chosen values are listed in Table~\ref{tab:sa_sweeps}. We verify at a representative point $(d, p) = (9, 0.10)$ that doubling the sweep counts in Table~\ref{tab:sa_sweeps} produces no statistically significant change in the logical error rate.

\begin{table}[htbp]
\caption{Number of SA sweeps per run as a function of code distance $d$ and noise model.}
\label{tab:sa_sweeps}
\renewcommand{\arraystretch}{1.05}
\setlength{\tabcolsep}{8pt}  
\begin{tabular}{ccccc}
\hline\hline
 & \multicolumn{2}{c}{Toric} & \multicolumn{2}{c}{Color} \\
$d$ & bit-flip & depol.\ & bit-flip & depol.\ \\
\hline
5  & 500  & 1000  & 500  & 1000  \\
7  & 800  & 1800  & 800  & 1500  \\
9  & 1000  & 2500  & 1000 & 2000 \\
11 & 1500 & 3500 & 1500 & 3000 \\
\hline\hline
\end{tabular}
\end{table}

We use $N=20$ class-restricted SA runs per logical class across all distances and noise models. The total trial budget per syndrome is therefore $N_{\mathrm{total}} = 20\,|\mathcal{L}|$, ranging from $40$ ($|\mathcal{L}|=2$, color bit-flip) to $320$ ($|\mathcal{L}|=16$, toric depolarizing).

\subsection{Random window decoder}
\label{sm:rw}

RW~\cite{dumer2017distance,pryadko2022qdistrnd} decoding on the Toric code under bit-flip noise generates candidate corrections as follows. A random column permutation is applied to the restricted classical parity-check matrix $H$ (representing the $Z$-type stabilizers) to obtain $H_{\mathrm{perm}}$, and the augmented system $[H_{\mathrm{perm}} \mid \boldsymbol{s}]$, with $\boldsymbol{s}$ the corresponding classical syndrome, is reduced to row echelon form to generate syndrome-consistent candidate solutions satisfying $H\boldsymbol{c}^T = \boldsymbol{s} \pmod 2$; the resulting candidates are retained. The stochasticity originates from the random column permutation, applied column-by-column ($g=1$ block size). Within a single trial, the procedure is repeated until $\nu_{\mathrm{win}}=5$ valid windows are accepted, with a maximum of $40$ retries per trial to handle rank-deficient instances.

RW operates under Strategy~A: every valid candidate from every accepted window is mapped to its logical class via the symplectic inner product with the logical observables and pooled into the corresponding $S_\ell$. AMLD applies the truncated free-energy decision [Eq.~\eqref{eq:DecisionRule}] on these pools; the MWD baseline operates on the same pools and returns the globally lowest-weight candidate. We use $N_{\mathrm{total}} = 100, 200, 300$ trials per syndrome at code distances $d = 4, 6, 8$, respectively.

\subsection{BP-OSD}
\label{sm:bposd}

We use the BP-OSD implementation in the \textsc{ldpc} Python package~\cite{ldpc_package,Roffe2020}. A single \texttt{BpOsdDecoder} instance is initialized with: BP method min-sum, maximum BP iterations $200$, min-sum scaling factor $0.625$, OSD method combination-sweep (\texttt{osd\_cs}), OSD order $24$, and channel probabilities set to the true physical error rate. The OSD order $24$ is chosen by sweeping $K \in \{8, 16, 24, 32\}$ and selecting the smallest value beyond which the logical error rate remains unchanged within a $1\sigma$ bootstrap error. This hyperparameter, like all other BP-OSD settings, is shared between the MWD baseline and AMLD.

Trial-dependent prior diversity is introduced by perturbing the per-qubit channel probabilities: at each trial $t \geq 2$, each qubit's channel probability is independently multiplied by a lognormal factor $\exp(\mathcal{N}(0, \sigma^2))$ with $\sigma = 0.25$ and clipped to $[10^{-15}, 1-10^{-15}]$. The first trial uses the unperturbed channel probabilities, ensuring the unperturbed BP-OSD output is always included in the candidate pool. For each syndrome, $N_{\mathrm{total}}$ independent trials are run and all syndrome-consistent corrections are mapped to their logical classes via the symplectic inner product with the logical observables (Strategy~A). We sweep $N_{\mathrm{total}}$ up to $200$ for the $[[144, 12, 12]]$ BB code at $p = 0.05$ (Fig.~\ref{fig:bposd_convergence}).

\section{Threshold evaluation}
\label{sm:threshold}

Logical error rates are estimated from $N_{\mathrm{shots}} = 2 \times 10^4$ independent noise realizations per data point, with bootstrap standard errors at $1\sigma$. Thresholds are extracted via the standard finite-size scaling ansatz
\begin{equation}
    p_L(p, d) = A + B x + C x^2, \qquad x = (p - p_{\mathrm{th}})\, d^{1/\nu},
    \label{eq:scaling}
\end{equation}
with $\{A, B, C, p_{\mathrm{th}}, \nu\}$ jointly fitted across all $(p, d)$ data points by nonlinear least squares as in Ref.~\cite{Wang2003}, weighting each point by its inverse bootstrap variance. Threshold uncertainties quoted in the main text are the $1\sigma$ standard errors derived from the fit covariance matrix.

\bibliography{references}

\end{document}